\documentclass[twocolumn,preprintnumbers,amsmath,amssymb,prl]{revtex4}

\usepackage{graphicx}
\usepackage{dcolumn}
\usepackage{bm}

\begin{document}
\title{Non-perturbative QED Model with Dressed States to Tackle HHG in Ultrashort
  Intense  Laser Pulses}
\author{Huayu Hu and Jianmin Yuan\footnote[2]{ Email:jmyuan@nudt.edu.cn}}
\address{Department of Physics,
National University of Defense Technology,
Changsha 410073, P. R. China}
\date{\today}

\begin{abstract}
A generalization of non-perturbative QED model for high harmonic
generation  is developed for the multi-mode optical field case. By
 introducing  classical-field-dressed quantized
Volkov states analytically, a formula to calculate HHG for hydrogen-like atom
in ultrashort intense laser pulse
is obtained, which has a simple intuitive interpretation. The
dressed state QED model indicates a new perspective to understand
HHG, for example, the presence of the weak even-order harmonic photons, which
has been verified by both theoretical analysis and numerical computation. 
Long
wavelength approximation 
and
strong field approximation are involved in the development of the formalism.
\end{abstract}
\maketitle

High order harmonic generation (HHG), that atoms subjected to an
intense laser field  emit more than a hundredth-order harmonic
photons, is the highest nonlinear optical phenomenon ever
observed\cite{experiments}. It has triggered a series of
experimental and theoretical researches. HHG not only  offers an
approach to the table-top source for coherent XUV and soft x-ray
generation\cite{xray}, but also to  attosecond light
pulses\cite{firatto}. The latter has been used to probe ultrafast
atomic and molecular dynamics, eg. atom ionization and correlation,
molecule rearrangement, molecule dissociation and so
on\cite{ionization}.

HHG has been widely understood as a three-step process --- the ionization,
propagation and recombination.
Calculations have been carried out by direct numerical integration of
the time-dependent Schr\"{o}dinger equation, using base expansion or grid
methods\cite{nolinearintegral}. In addition, various theoretical models
concerning essential physical processes in intense laser field and less computer
intensive
are widely studied. 
For example, the KFR model for the scattering\cite{KFR}, the ADK model for the ionization\cite{ADK}, the Lewenstein
model for HHG\cite{SFA}, and so on. These models are based on the
classical treatment of the laser field. In these semiclassical models, the
HHG spectrum is obtained by the Fourier transform of the
time-dependent electric dipole, or the dipole acceleration.

On the other hand, the observed discrete HHG spectrum indicates
that it is an optical-conversion process in which $q$ pump photons
are combined  
to produce a single, harmonic photon\cite{jgaoprl}.  This picture favors the
QED approach which treats the photon field as a quantized field. It is capable of presenting a clearer picture of
how the photons are being absorbed and emitted, and how the energy
and momentum transfer between the optical field and the atom.
A  non-perturbative QED scattering theory of single-mode case
has been developed for HHG\cite{guo}. It provides a new perspective to understand
HHG,  for example, rather than by extracting HHG spectrum from the
time-dependent atomic wavepacket,      it self-contains harmonic photon
emitting  in the theory.  Similar phenomena at much higher intensities can be
calculated on its framework. It has
successfully reproduced the salient characteristics of experimental spectra
and revealed the connection between high harmonic generation and above
threshold ionization\cite{guo,jgaoprl,korean}.

The essential part of QED model for HHG in a intense laser field is the transition matrix
element\cite{guo,jgaoprl,korean}
\begin{equation}\label{eq:1}
T_{fi}=\sum_{\mu}\langle\Psi_{f}|V^{'}|\Psi_{\mu}\rangle\langle\Psi_{\mu}|V|\Psi_{i}\rangle\delta_{\mu}\;,
\end{equation}
where $\Psi_{i}$ is the initial state of the atom-photon system,
$\Psi_{f}$ is the final state with a harmonic photon, $\Psi_{\mu}$
is a  quantized-field Volkov state, and $V\,(V^{'})$ is the
interaction potential. The symbol $\delta_{\mu}$ represents energy
conservation condition in the  non-relativistic case.
Eq.~(\ref{eq:1})    gives a physical picture that the atom in an
intense optical field can simultaneously absorb more than one photon
to transit to the quantized-field Volkov states, and finally emit a
harmonic photon to transit to the final state.

The quantized-field Volkov states are the pivotal part of the QED
approach. Classical Volkov states\cite{volkov}, which are the solutions of
time-dependent Schr\"{o}dinger equation for a free electron in a
classical electromagnetic field, are incorporated in the KFR
theory to form a non-perturbative scattering framework for
multiphoton ionization (MPI). Later, Guo, {\AA}berg and Drake
obtained analytical solutions for an free electron in an intense
quantized optical field\cite{Aberg,Drake}---the  quantized-field
Volkov states, based on which the QED calculation for  HHG is
developed.

Most of the calculations for HHG in QED have been carried out by
assuming  the fundamental laser field to be single-mode
\cite{guo,jgaoprl,korean}.  The formal scattering theory, which takes
the interaction time as infinity, is always used. This is a good
approximation to many experiments in which the laser pulse's
duration is dozens of optical cycles long so that the laser field
can be considered as single-mode. However, it still limits the usage
of the method  in certain ways. For example, the effect of the laser
pulse's shape cannot be studied,
 the calculated HHG spectrum only
contains the photon frequencies which are multiples of the
fundamental frequency, and ultrafast atom-laser interactions are hard to
describe.  



\begin{figure}
\includegraphics[height=2in,width=3.3in]{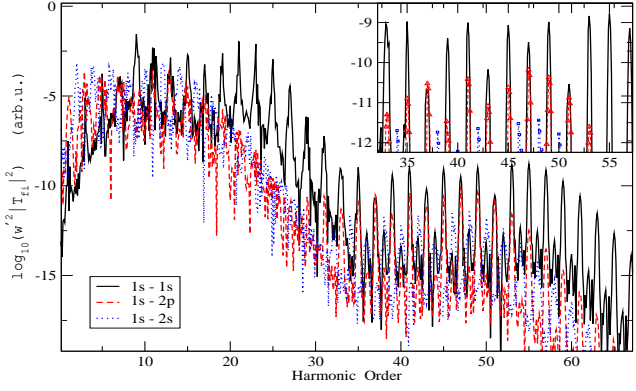}
\caption{\label{fig1} HHG spectrum of H atom  for the laser
parameters: $\lambda=0.8\mu m$,  peak intensity
$I=3.5\times10^{14}W/cm^{2}$, gaussian pulse shape with $FWHM=16
fs$.
 The solid line is $1s-Volkov\, state-1s$ transition, the dash line is
 $1s-Volkov\, state-2p$ transition, and the dot line is  $1s-Volkov\, state-2s$
 transition. The inset shows a part of the spectrum from harmonic order 33
to 55.}
\end{figure}
In the present study, we improve the QED model to the multi-mode case by taking the
Born-Oppemheimer approximation,  that  the photon
mode is a far more  rapidly varying degree of freedom than the electron, so
the stimulated electron-photon interactions dress the atomic electron.
Therefore, in the multi-mode case, the
interaction is between the photons  and the
photon-cloud-dressed  electron, instead of the naked electron as in the
previous models. Similar idea has been used in studying laser-assisted bremsstrahlung
in highly intense laser field\cite{keitel}.


Two key techniques are used in our theory. Firstly, we treat the strongest
mode of the laser field as the quantized mode, which has the same optical cycle
as the laser pulse. By deriving the classical-field-dressed
quantized Volkov states, the effects of all the other modes are incorporated as
a classical distortion field which dresses the quantized Volkov state. Secondly,
we make use of  the Fourier transform of the evolution operator,
instead of the time-independent formal scattering formalism. This allows us to
include time in the theory. The inverse operators are transformed to
computable integrals by the equations we derive.

We differ from the previous studies\cite{guo,jgaoprl,korean} by using coherent states instead of Fock
states to describe the laser field. Coherent state
$|\alpha_{1}\alpha_{2}\cdots\rangle$, where
$|\alpha_{k}\rangle
=\exp^{-\frac{|\alpha_{k}|^{2}}{2}}\sum\limits_{n=0}\limits^{\infty}{\frac{\alpha_{k}^{n}}{\sqrt{n!}}|n\rangle}$
is the eigenstate of the  annihilation operator $\hat{a}_{k}$ with
$\hat{a}_{k}|\alpha_{k}\rangle=\alpha_{k}|\alpha_{k}\rangle$, is a better
approximation to the real laser field\cite{coherent}:
$\langle
\alpha_{1}(t)\alpha_{2}(t)\cdots|\mathbf{\hat{A}}|\alpha_{1}(t)\alpha_{2}(t)\cdots
\rangle
 =\sum{\mathbf{A}_{i}(t)}=\mathbf{A}(t)$, where $\mathbf{A}(t)=\hat\epsilon A(t)$ is the electromagnetic field vector.
 In this treatment, the absolute value and the phase of $\alpha_{k}$ are in accordance
 with the  amplitude and the phase of the kth mode of the laser field.
 Therefore it provides a way  to characterize the time-dependence of the
 laser pulse by $\alpha_{k}=\frac{1}{T}\int\limits_{0}\limits^{T}d\tau
 e^{iw_{k}\tau}A(\tau).
$
The long wavelength limit is taken in the present scheme. Walser et
al.\cite{multipole} have considered the multipole contributions to HHG, and
found such effects be small below an intensity of
about $10^{17}  W/cm^{2}$.

In the non-relativistic case, the total  hamiltonian of a
hydrogen-like atom in the optical field is
 $  \hat{H}=\hat{H_{0}}+\hat{V}_{T}$
, where
  $\hat{H_{0}}=\frac{1}{2m}\mathbf{\hat{p}}^{2}+
\sum{\omega_{k}(\hat{N}_{k}+\frac{1}{2})}+\omega^{'}(\hat{N}^{'}+\frac{1}{2})+\hat{U}$
is the noninteraction part of the Hamiltonian. $\hat{U}$ is the
atomic binding potential; $\hat{N^{'}}$ and $\hat{N}_{k}$ are
photon number operators of harmonic mode with frequency
$\omega^{'}$ and  the kth laser mode with frequency $\omega_{k}$,
respectively. $\hat{V}_{T}$ is the electron-photon interaction term
$\hat{V}_{T}=-\frac{e}{m}\mathbf{\hat{p}}\cdot\mathbf{\hat{A}}+\frac{e^{2}}{2m}\mathbf{\hat{A}}^{2}$. 
Due to the weakness of the harmonic mode, $\hat{V}_{T}$ can be divided as
$\hat{V}_{T}=\hat{V}+\hat{V}^{'}$
\begin{eqnarray}
&&\hat{V}=-\frac{e}{m}(\mathbf{\hat{A}}_c+\mathbf{\hat{A}}_{s})\cdot
\mathbf{\hat{p}}+\frac{e^{2}}{2m}(\mathbf{\hat{A}}_c+{\mathbf{\hat{A}}_{s}})^{2}\;,\\
&&\hat{V}^{'}=-\frac{e}{m} \mathbf{\hat{A}}^{'}\cdot\mathbf{\hat{p}}+\frac{e^{2}}{m}\mathbf{\hat{A}}^{'}
\cdot (\mathbf{\hat{A}}_c+\mathbf{\hat{A}}_{s})\;.
\end{eqnarray}
For simplicity, we denote $\mathbf{\hat{A}}^{'}$ as the harmonic mode,
$\mathbf{\hat{A}}_c$ as the quantized laser mode,
and $\mathbf{\hat{A}}_s=\sum\limits_{s\neq c}\mathbf{\hat{A}}_s $ as all the
other laser modes, so $\mathbf{\hat{A}}=\mathbf{\hat{A}}_c+\mathbf{\hat{A}}_s+\mathbf{\hat{A}}^{'}$.
In general, $\mathbf{\hat{A}}_u=g_{u} (\hat\epsilon_{u} \hat{a}_{u} e^{i\mathbf{k}_{u}\cdot\mathbf{r}}+c.c.)$,
where the polarization vector $\hat{\epsilon}$ is given by
$\hat{\epsilon}=cos(\xi/2)\hat{\epsilon}_{x}+isin(\xi/2)\hat{\epsilon}_{y}$,
where $\xi$ is a measure of the degree of ellipticity, then $\xi=0$
and $\pi/2$ corresponding to linear and circular polarization,
respectively \cite{guo}. The driving optical field in our
 calculation is linearly polarized.

The non-perturbative approach for HHG is based on the fourier
transform of the evolution operator $\hat{U}(t,t^{'})$
\cite{evolution}
\begin{align}\label{fourier}
\hat{U}(t,t^{'})=& -\frac{1}{2\pi
  i}\int_{-\infty}^{+\infty}dz\frac{e^{-iz\tau}}{z-\hat{H}+i\epsilon} \Theta(\tau) \nonumber \\
& +\frac{1}{2\pi i}\int_{-\infty}^{+\infty}dz\frac{e^{-iz\tau}}{z-\hat{H}-i\epsilon}
\Theta(-\tau)\;,
\end{align}
where $\tau=t-t^{'}$, $\Theta(\tau)=1$ when $\tau\geq 0$ and  $\Theta(\tau)=0$
when $\tau < 0$ ; $\epsilon\rightarrow 0_{+}$. The operator can be expanded, giving
the matrix element for the  harmonic generation process as
\cite{guo,jgaoprl,korean}
\begin{align}\label{eq:6}
T_{fi}&=\langle\Psi_{f}(T)|\hat{U}(T,0)|\Psi_{i}(0)\rangle\nonumber\\
&=-\frac{1}{2\pi i}\int_{-\infty}^{+\infty} dz
e^{-izT}e^{i(E^{a}_{f}+\frac{3}{2}w^{'})T}
\times\langle\Phi_{f},\alpha_{f}(T),1^{'}|\nonumber\\
&
\times\frac{1}{z-\hat{H}_{0}+i\epsilon}V^{'}\frac{1}{z-\hat{H}+i\epsilon}V\frac{1}{z-\hat{H}_{0}+i\epsilon}\nonumber\\
&\times|\Phi_{i},
\alpha_{i}(0),0^{'}\rangle \;,
\end{align}
where the system's initial and final states are decoupled states of the
electron and the laser field, and the laser pulse is
confined to the interval $[0,T]$.  $\Phi_{i,f}$
is the  wave function of the atomic electron with binding energy
$E^{a}_{i,f}$; $|n^{'}\rangle$, with $n=0,1$, is
the harmonic photon Fock state, and
$|\alpha_{i,f}(t)\rangle=e^{-i\frac{w_{1}+w_{2}+\cdots}{2}t}|\alpha^{1}_{i,f}e^{-iw_{1}t},\alpha^{2}_{i,f}
e^{-iw_{2}t},\cdots\rangle$ is the coherent laser state .


If $\Phi$ is the eigenstate of the atomic hamiltonian and $\hat{H}_{0}$ is the
noninteraction hamiltonian, we get the relations
\begin{eqnarray}\label{rel}
&&\frac{1}{z-\hat{H}_{0}+i\epsilon}|\Phi(t_{0})\alpha(t_{0})\rangle \nonumber\\
=&&
 -i\int^{+\infty}_{-\infty} dt e^{izt}|\Phi(t+t_{0})\alpha(t+t_{0})\rangle \Theta(t)\nonumber\\
=&& -i\int^{+\infty}_{0} dt e^{izt}|\Phi(t+t_{0})\alpha(t+t_{0})\rangle \;, \\
&&\langle \Phi(t_{0})\alpha(t_{0})|\frac{1}{z-\hat{H}_{0}+i\epsilon} \nonumber\\
=&&
 -i\int^{+\infty}_{-\infty} dt e^{-izt} \langle \Phi(t+t_{0})\alpha(t+t_{0})| \Theta(-t)\nonumber\\
=&& -i\int^{0}_{-\infty} dt e^{-izt} \langle \Phi(t+t_{0})\alpha(t+t_{0})|\;.
\end{eqnarray}
Then the transition matrix $T_{fi}$ becomes 
\begin{align}\label{formore}
T_{fi}&=\frac{1}{2\pi i}\int_{-\infty}^{+\infty} dz\int^{0}_{-\infty}dt_{1}\int^{+\infty}_{0}
dt_{2}    e^{-i[z-(E^{a}_{f}+\frac{3}{2}w^{'})](T+t_{1})}\nonumber\\
 &\times \langle\Phi_{f},\alpha_{f}
(T+t_{1}),1^{'}|
V^{'}\frac{1}{z-\hat{H}+i\epsilon}V|\Phi_{i},\alpha_{i}(t_{2}),0^{'}\rangle\nonumber\\
&\times e^{-i E^{a}_{i}t_{2}}\;.
\end{align}
Therefore, the  two inverse operators have been
transformed into computable integrals.

To apply this strategy further to tackle Eq. (\ref{formore}), we
need the knowledge of the eigenstates of $\hat{H}$. Due to the
weakness of $\hat{V^{'}}$ and the strong field approximation(SFA),
the binding potential $\hat{U}$ can be neglected for continuum
electron states\cite{SFA}, hence only the leading term of the
inverse operator $\frac{1}{z-\hat{H}+i\epsilon}$ has to be kept,
thus enabling $\hat{U}$ and $\hat{V}^{'}$ in the denominator in Eq.
(\ref{formore}) to be dropped\cite{guo}.

In the one-mode case, the eigenstates of such a Hamiltonian 
are the quantized-field Volkov
states\cite{Aberg}: $
 \Psi_{pn}=V^{-1/2}_{e}\sum\limits_{j=-n}\limits^{\infty}e^{i[\mathbf{p}+
 (u_{p}-j)\mathbf{k}]\cdot \mathbf{r}}
 \times\mathcal{J}_{j}
(\zeta,\eta,\phi_{\xi})^{*}\times e^{-ij\phi_{\xi}}|n+j\rangle
$
with corresponding energy eigenvalues
$E_{pn}=\frac{\mathbf{p}^{2}}{2 m}+(n+1/2+u_{p})w $. 
$u_{p}= \frac{e^{2}\Lambda^{2}}{mw}$ is the ponderomotive energy in
units of the photon energy of the laser, where $\Lambda$, as the limit of
$g\sqrt{n} (g\rightarrow 0, \sqrt{n}\rightarrow \infty)$, is the half amplitude
of the classical
field; $\zeta=\frac{2e\Lambda}{mw}|\mathbf{p}\cdot\hat{\epsilon}|$ ,
$\eta=\frac{1}{2}u_{p}cos\xi$ and $\phi_{\xi}=\arctan(\frac{p_{y}}{p_{x}}\tan\frac{\xi}{2})$.
The  generalized Bessel function $\mathcal{J}_{j}(\zeta,\eta,\phi_{\xi})$
  is
$\mathcal{J}_{j}(\zeta,\eta,\phi_{\xi})=\sum\limits_{m=-\infty}^{\infty}J_{-j-2m}(\zeta)
J_{m}(\eta)e^{i2m\phi_{\xi}}$.

In the  multi-mode case, with one specified quantized mode ($A_{c}$), we obtain
the classical-field-dressed quantized-field Volkov states
\begin{align}\label{eq:2}
\widetilde{\Psi}^{c}_{pn}(t)&=e^{ie\sum\mathbf{A}_{s}(t)\cdot\mathbf{r}}| \Psi^{c}_{pn}(t)\rangle
|\alpha^{s}(t)\rangle,\nonumber\\
&=V_{e}^{-1/2}e^{ie\sum\mathbf{A}_{s}(t)\cdot\mathbf{r}}|\alpha^{s}(t)\rangle
\sum\limits_{j=-n_{c}}\limits^{\infty}e^{i[\mathbf{p}+
 (u^{c}_{p}-j)\mathbf{k_{c}}]\cdot \mathbf{r}}\nonumber\\
 &\times\mathcal{J}_{j}
(\zeta_{c},\eta_{c},\phi^{c}_{\xi})^{*} e^{-ij\phi^{c}_{\xi}}|n_{c}+j\rangle e^{-iE_{pn_{c}}t}\;.
\end{align}
As before, the script $s$ represents all the other modes in the laser field
except the mode distinguished by $c$.
In the large-photon-number limit, where $a^{+}|\alpha\rangle=\alpha^{*}|\alpha\rangle $ for $|\alpha|\rightarrow
\infty$, $\widetilde{\Psi}^{c}_{pn}$  has the property
 \begin{align}\label{eq:5}
& (\frac{1}{2m}(\mathbf{\hat p}-e\mathbf{\hat
  A})^{2}+\sum{\omega_{k}(\hat{N}_{k}+\frac{1}{2})}) \widetilde{\Psi}^{c}_{pn}\nonumber\\
  =&(E^{c}_{pn}+E_{s})\widetilde{\Psi}^{c}_{pn}
 \end{align}
Where $E_{s}=\sum w_{s}|\alpha_{s}|^{2}$.  Eq. (\ref{eq:5}) verifies that  the
classical-field-dressed  quantized-field Volkov state is the
 approximate
 eigenstate of the  electron-laser subsystem in the multi-mode field. Therefore,
a relation similar to Eq. (\ref{rel}) can be obtained
\begin{align}\label{eq:4}
&\frac{1}{z-H+i\epsilon} e^{ie\sum\mathbf{A}^{s}(t_{0})\cdot\mathbf{r}}
 |\Psi^{c}_{pn}(t_{0})\rangle|\alpha^{s}(t_{0})\rangle\nonumber\\
=& -i\int^{+\infty}_{0} dt e^{izt}e^{ie\sum\mathbf{A}^{s}(t_{0}+t)\cdot\mathbf{r}}\nonumber\\
&\times |\Psi^{c}_{pn}(t_{0}+t)\rangle|\alpha^{s}(t_{0}+t)\rangle\;.
\end{align}

Combining Eq. (\ref{formore}) and (\ref{eq:4}), and after the
integration of the energy $z$, we get
\begin{widetext}
\begin{align}\label{tranf}
T_{fi}
=&\frac{1}{2\pi}\sum\limits_{pn} \int^{T}_{0} d\tau_{1}
\int^{T}_{\tau_{1}}d\tau_{2}\langle\Phi_{f},\alpha^{c}_{f} ,1^{'}|
V^{'}(\tau_{2})e^{ie\sum\mathbf{A}^{s}(\tau_{2})\cdot\mathbf{r}}
 |\Psi^{c}_{pn},0^{'}\rangle \nonumber\\
&\times e^{i(E^{a}+\omega^{'}_{f}-E^{c}_{pn})\tau_{2}}
e^{-i(E^{a}_{i}-E^{c}_{pn})\tau_{1}} \langle \Psi^{c}_{pn},0^{'}
|e^{-ie\sum\mathbf{A}^{s}(\tau_{1}) \cdot \mathbf{r}}
V(\tau_{1})|\Phi_{i},\alpha^{c}_{i},0^{'}\rangle \;,
\end{align}
\end{widetext}
where
$V(\tau_{1})=\frac{e^{2}}{2m}\mathbf{A}^2(\tau_{1})-\frac{e}{m}\mathbf{A}(\tau_{1})\cdot(\mathbf{p}+e
\sum\mathbf{A}^{s}
(\tau_{1}) +(u^{c}_{p}-j)\mathbf{k}_{c})$ and
 $V'(\tau_{2})=\frac{e^{2}}{2m}g'\hat{\epsilon}^{'*}\cdot\mathbf{A}(\tau_{2})-\frac{e}{m}
 g'\hat{\epsilon}^{'*}\cdot(\mathbf{p}+e\sum\mathbf{A}^{s}(\tau_{2})
+(u^{c}_{p}-j)\mathbf{k}_{c}$.

Eq.~(\ref{tranf}) indicates  that the probability amplitude of
emitting a harmonic photon is the sum of the probability amplitudes
of all the individual  events in which the electron is ionized at
some time and recombines with the nucleus to emit the photon at a
later time. The picture is intuitive, but is not trivial for these
probabilities add up in an interference way.

\begin{figure}
\includegraphics[height=2in,width=3.2in]{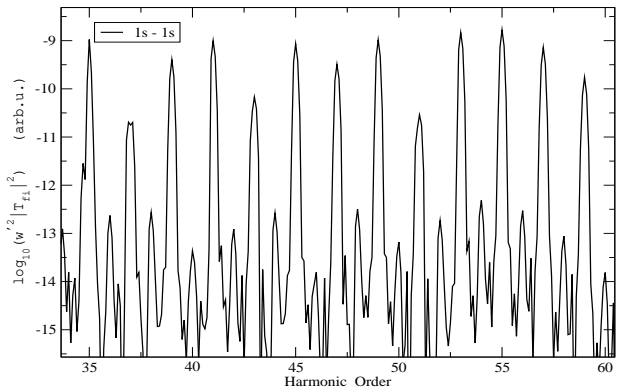}
\caption{\label{fig2} One Part of HHG spectrum of H atom for the same laser
  parameters as in
  Fig. (\ref{fig1}). It only shows HHG from $1s-Volkov\, state-1s$ transition. }
\end{figure}

HHG of H atom in an intense ultrashort laser pulse has been calculated as a numerical example. The laser parameters
are:
$\lambda=0.8\mu m$,  peak intensity $I=3.5\times10^{14}W/cm^{2}$, gaussian pulse
shape with $FWHM=16 fs$, that is approximately 6 optical cycles.
The atom is assumed to be initially in the ground
state. Different lines in  Fig (\ref{fig1}) represent  different final state
cases, here the $1s$, $2p$, $2s$ states respectively. It shows that, among the
three transitions considered here,
$1s-Volkov\, states-1s$
transition dominates the harmonic photon generation.
The contributions from
other transitions are about $10^{-1\sim-2}$ smaller in the plateau.
It demonstrates that the SFA, where only the recombination to the ground state is taken into account,  is a good
approximation in this case.
This QED model can then be used to complement
other models by indicating what bound states should be included in the
approximation.

The fine structures of HHG in Fig (\ref{fig2}) show the generation
of even-order harmonic photons. The presence of even harmonics can
be explained as the contribution of
$\frac{e^{2}}{2m}\mathbf{\hat{A}}^{2}$, which can not be considered
as a perturbation as in the weak light field case. Transition
through $-\frac{e}{m}\mathbf{\hat{p}}\cdot\mathbf{\hat{A}}$ changes
 the atomic state's parity, while transition through
$\frac{e^{2}}{2m}\mathbf{\hat{A}}^{2}$ maintains it. Therefore, the argument for
the absence of even harmonics, that is the selection rule for the dipole transition, does
not hold here.
 In this laser parameter condition,
the intensity of the even harmonics is much weaker than the odd
ones, with a factor $10^{-4}$. This is in accordance with the
experiments where  only odd harmonics are observed. However, in the
semi-classical model, or numerically solving TDSE, the information of even
harmonics is often buried by the numerical noise or more importantly, by other
weaker transitions,
like $1s-Volkov\, states-2s$ and $1s-Volkov\, states-2p$  shown
here.

In conclusion, a generalization of non-perturbative QED model for
HHG has been developed to the multi-mode case. In our QED model,
HHG produced by ultrashort laser pulse can be calculated for
hydrogen-like atoms. A numerical example has been calculated for H
atom in intense ultrashort laser pulse and it reveals a clear
evidence of even harmonics generation.




\begin{acknowledgments}
This work was supported by the National Natural Science Foundation
of China under Grant No. 10734140, the National Basic Research
Program of China (973 Program) under Grant No. 2007CB815105, and
the National High-Tech ICF Committee in China. We would also like to
thank Dr Zengxiu Zhao for helpful discussions.
\end{acknowledgments}

\end{document}